\documentclass[%
 amsmath,amssymb,
 reprint,%
]{revtex4-1}

\usepackage{graphicx}
\usepackage{dcolumn}
\usepackage{bm}

\usepackage[utf8]{inputenc}
\usepackage[T1]{fontenc}
\usepackage{mathptmx}
\usepackage{etoolbox}
\usepackage{xcolor}

\definecolor{lgray}{gray}{0.9}

\usepackage{color}

\begin{document}
\title{Integrated optical quantum memory controlled by electro-optic effect}

\author{Xia-Xia Li}
\affiliation{Key Laboratory of advanced optoelectronic quantum architecture and measurements of Ministry of Education, Beijing Key Laboratory of Nanophotonics \& Ultrafine Optoelectronic Systems, School of Physics, Beijing Institute of Technology, 100081, Beijing, China}
\author{Pai Zhou}
\affiliation{Key Laboratory of advanced optoelectronic quantum architecture and measurements of Ministry of Education, Beijing Key Laboratory of Nanophotonics \& Ultrafine Optoelectronic Systems, School of Physics, Beijing Institute of Technology, 100081, Beijing, China}

\author{Yu-Hui Chen}\email{stephen.chen@bit.edu.cn}
\affiliation{Key Laboratory of advanced optoelectronic quantum architecture and measurements of Ministry of Education, Beijing Key Laboratory of Nanophotonics \& Ultrafine Optoelectronic Systems, School of Physics, Beijing Institute of Technology, 100081, Beijing, China}
\author{Xiangdong Zhang}
\affiliation{Key Laboratory of advanced optoelectronic quantum architecture and measurements of Ministry of Education, Beijing Key Laboratory of Nanophotonics \& Ultrafine Optoelectronic Systems, School of Physics, Beijing Institute of Technology, 100081, Beijing, China}

\date{\today}

\begin{abstract}
Integrated optical quantum memories are a scalable solution to synchronize a large number of quantum nodes. Without compact quantum memories, some astonishing quantum applications such as distributed quantum computing and quantum sensor networks would not be possible. Rather than find a specific material that meets all the requirements of an on-chip quantum memory as other protocols usually do, we propose to assign the memory requirements on coherent storage and controllability to rare earth ions and a lithium niobate crystal, respectively. Specifically, optical quantum states are stored in an erbium-doped lithium niobate micro-cavity by utilizing the electro-optic effect of lithium niobate.  The cavity frequency can be shifted by an external electric field, thus modifying the resonance condition between the cavity and the collective atomic excitation. This effect is further used to suppress or enhance the emission of photon echoes. Our calculated results show that high efficiency and low noise performance is achievable.  
\end{abstract}

\maketitle
Recent developments in quantum information have placed practical quantum computers just over the horizon. To fully unlock the potential of quantum computing, it is desired to connect multiple quantum nodes together to build a quantum internet\cite{Wehner2018,Simon2017,Bush2021}, which in turn could enable some astonishing applications such as distributed quantum computing\cite{VanMeter2016} and quantum sensor networks \cite{Eldredge2018,Zhong2019}. Such a quantum internet needs quantum memories to store quantum states for on-demand reading \cite{Lvovsky2009}. Otherwise, distant quantum qubits can not be synchronized and any failure in any node would require the whole system to restart from the beginning, making a large quantum internet almost impossible. Similar to a quantum computer that needs many qubits to demonstrate its quantum advances \cite{Arute2019},  accessing the full potential of quantum networks requires scaling of multiple nodes in global distances\cite{Reiserer2022}. In this aspect, integrating multiple units into a single chip appears to be the most effective approach for improving the scalability of a memory system, as evidenced by the success of modern integrated circuits and integrated photonic chips. Unlike the rapid advances in quantum processor chips, while quantum memories of $\mu$m size have been fabricated \cite{Zhong2019,Liu2020,Craiciu2021,Tian2017}, an on-chip quantum memory with good performance metrics, and at the same time showing scalability with other photonic devices, remains an outstanding challenge.   

Due to their solid-state nature and long coherence times\cite{Thiel2011}, rare-earth doped crystals are one of the best candidates for building on-chip quantum memories \cite{Zhong2019}. Though there have been impressive attempts to build memories by patterning nanostructures on rare-earth-doped crystals \cite{Zhong2017,2015nano,Liu20,2010broadband}, achieving faithful performance is hampered by the difficulties imposed by the small size of a chip. For example, it is desirable for an atomic frequency comb (AFC) \cite{2020on} memory to have large optical absorption such that it can have a high storage efficiency \cite{Liu20}. However, for a memory based on a waveguide architecture, its length on a chip limits the achievable optical depth. Increasing the ion density seems a straightforward solution, but at the cost of coherence times \cite{Chen2021b}. Besides, quantum memory protocols such as controlled reversible inhomogeneous broadening (CRIB) \cite{2008Analytic,2012Spatial} rely on the assistance of applied magnetic or electric fields. The need for a magnetic (or electric) field with a large gradient on a chip not only complicates the architecture, but also puts additional constraints on the magnetic or electric properties of the working ions. As a result, current on-chip memories are much less competitive {in} performance {than} their bulk counterparts \cite{Ho2016,Dajczgewand14}. When designing a practical on-chip quantum memory, both the requirements for faithful performance and easy scalability should be taken into account.


Lithium-niobate (LiNbO$_{3}$) is a material with the name of photonic silicon, being famous for its large electro-, nonlinear-, and acousto-optic coefficients. Photonic devices based on LiNbO$_{3}$ have rapidly advanced with the commercialization of lithium-niobate-on-insulators  thin film. For example, the development of LiNbO$_{3}$ waveguides \cite{2020Incorporation}, high-quality micro-resonators \cite{2017On}, high-frequency modulators \cite{2020Lithium}, and on-chip telecom lasers \cite{Luo2021} have made thin-film LiNbO$_{3}$ an unprecedented platform for integrated photonic circuits for both classical and quantum applications. 

Rather than imposing excessive requirements {on} the medium that functions as an integrated memory, we propose a scheme based on erbium doped lithium niobate (Er$^{3+}$:LiNbO$_3$) to realize optical quantum memories by using the electro-optic effect of LiNbO$_{3}$. The scheme relies on an electric-optically tuned cavity to control the emission of photon echoes, resulting in high storage efficiency and low noise performance. Using cavity detunings to silence the echo has been investigated in the microwave regime \cite{Afzelius2013,Julsgaard2013}, the first demonstration of which has recently been reported \cite{Ranjan2022}. But its efficiency still needs to be improved for practical quantum applications. Our memory scheme inherits some obvious industrial advantages of thin-film Er$^{3+}$:LiNbO$_3$, such as the fabrication maturity of LiNbO$_{3}$ thin films and the 1.5\,$\mu$m emission of erbium ions, being easy to scale up with multiple independent memory units in one chip.

Similar to photon-echo quantum memories\cite{Moiseev2004,2011Revival, 2011Photon,Afzelius2013,Julsgaard2013, 2016light, Ma2021a}, our approach is modified from the two-pulse echo scheme. The two-pulse echo is not suitable for optical quantum storage, because the echo is emitted from a population-inverted medium. The gain and spontaneous emission of the medium ruins the fidelity of echo \cite{Moiseev2004,2008Why}. Applying a second $\pi$ pulse brings ions back to their ground state; therefore the secondary echo can be used as a readout of the stored quantum state. However,  the echo rephased by the first $\pi$ pulse already emits a large part of the stored energy, leading to efficiency reduction of the second echo. To use the second echo as a memory readout, it is thus necessary to eliminate the emission of the first one. In the following, we detail how to use the electro-optic effect of LiNbO$_{3}$ to silence the undesired echo.

\begin{figure}[h]
\centering
\includegraphics[width=0.45\textwidth]{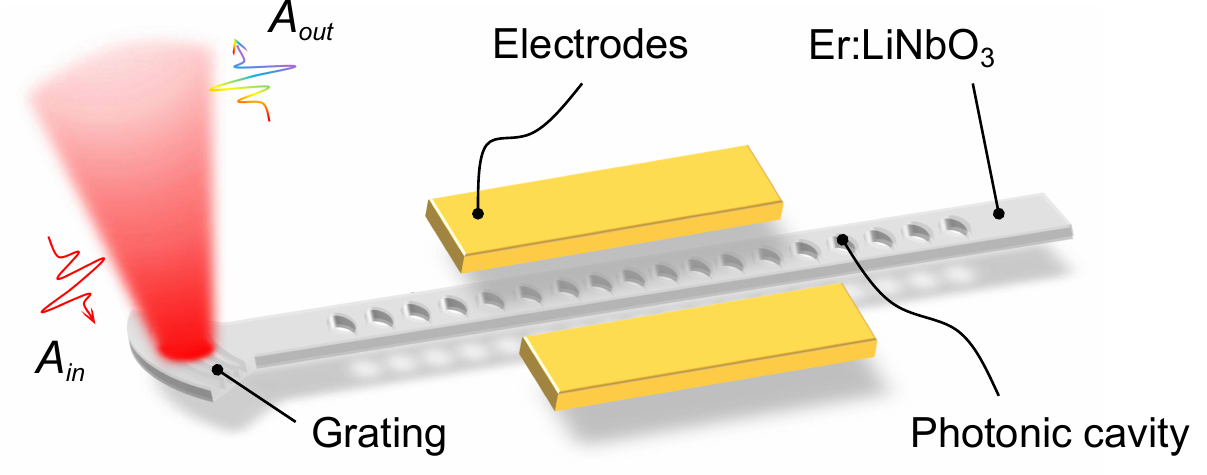}
\caption{Schematic of the cavity-controlled echo scheme. A photonic-crystal  nano-beam cavity is fabricated on an Er$^{3+}$:LiNbO$_3$ thin film. A grating coupler enables light input and output of the cavity. Electrodes located on two sides of the cavity are used to shift the cavity frequency.  }
\label{fig.idea}
\end{figure}

\begin{figure}[h]
\centering
\includegraphics[width=0.45\textwidth]{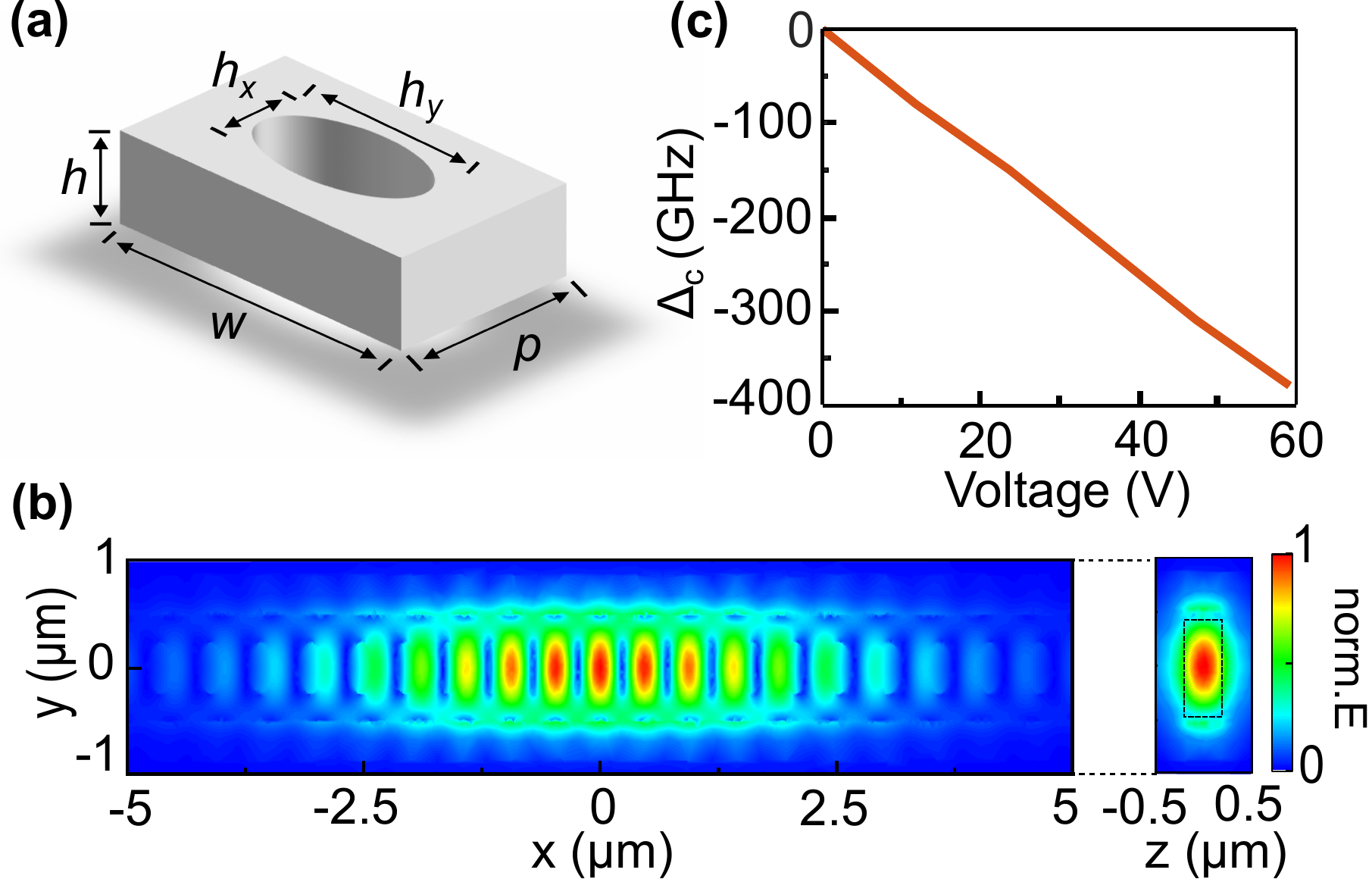}
\caption{Simulation of the nano-beam cavity. (a) The structure of a periodic unit of the nano-beam cavity. (b) Simulated electric field distribution of the nano-beam cavity. Left, $x-y$ plane from {the} top view; right, $y-z$ plane from {the} side view. (c) The cavity frequency shift $\Delta_c$ as a function of applied voltage, where the space distance between two electrodes is  4\,$\mu$m. } 
\label{fig.opt_cavity}
\end{figure}

The setup that we consider is shown diagrammatically in Fig. \ref{fig.idea}. A photonic-crystal nano-beam cavity is fabricated in {an} Er$^{3+}$:LiNbO$_3$ thin film. Light is input into (and output from) a one-sided cavity through a grating coupler. Two electrodes are used to supply electric field to utilize the electro-optic effect of LiNbO$_{3}$. The width and the thickness of the cavity are $w=900$\,nm and $h=390$\,nm, respectively. A photonic bandgap around 1.5 $\mu$m is obtained by milling oval holes with {$h_x=220$\,nm, $h_y=490$\,nm,} and a period of $p=530$\,nm, as illustrated in Fig. \ref{fig.opt_cavity}(a). A cavity resonance is formed by changing the period of the five central holes to $460$\,nm. Figure \ref{fig.opt_cavity}(b) shows the simulated electric-field distribution of the cavity at the resonant wavelength of 1.5 $\mu$m. 

The effective refractive index of LiNbO$_{3}$ in our simulations is modelled as  $n_{\text{eff}}=n_{0}-\frac{1}{2}n{_0}^3 r E$, where $n_{0} = 2.21$  is the refractive index of LiNbO$_{3}$ without applied voltage, $r = 30.9\times10^{-12}m/V$ is a constant related to the electro-optic coefficient of LiNbO$_{3}$ \cite{Pan:19}, and $E$ is the applied DC electric field. According to our simulations, the cavity frequency $\Delta_c$  shifts linearly with the change of refractive-index: $\Delta_c \propto \Delta n_{\text{eff}}$, as shown in Fig.2(c).  Applying a 10\,V voltage to the structure can shift the frequency $\Delta_c$  {by} approximately 60\,GHz. 
 
For a collection of $N$ two-level erbium ions interacting with such an optical cavity, the dynamic equations for an ion are: 
\begin{equation} \label{eq.rho12}
\begin{split}
\dot{\rho}_{21}(\mathbf{r}) & =(-i\delta-\frac {1} {T_2})\rho_{21}+ig(\mathbf{r}) a (\rho_{11}-\rho_{22}),\\
\dot{\rho}_{22}(\mathbf{r}) & =-\frac {1} {T_1}\rho_{22}+ig(\mathbf{r}) a \rho_{12}-ig(\mathbf{r}) a^\dagger \rho_{21},\\
\end{split}
\end{equation}
where $\rho_{ij}$ are the elements of the two-level density matrix $\rho$ with $i=1,2$, indicating the ground and excited state respectively, $a$ is the cavity-mode operator, $\delta$ is the atomic frequencies, $T_1$ is the excited-state lifetime, $T_2$ is the optical coherence time, and $g(\mathbf{r})$ denotes the single-photon Rabi frequency. It is assumed in the above equations that the cavity and atomic operators are uncorrelated, which is a good approximation even in the regime of single excitation. The inhomogeneous atomic detuning $\delta$ is subject to a Gaussian distribution $G(\delta - \delta_0 )$,  where $\delta_0$ is the center of the inhomogeneous line, and $\int G(\delta- \delta_0) d \delta =1$. The ensemble polarization of erbium ions is
\begin{equation} \label{eq.polar}
    P = N \int  \int_{-\infty}^{+\infty}\rho_{21}(\mathbf{r}, \delta)G(\delta-\delta_{0})d\delta d\mathbf{r},
\end{equation}
where $N$ is the total number of erbium ions. 

Different from plane wave propagating along a crystal, when light is input into a cavity \cite{Longdell2008, Patrick2010}, the time that the optical field needs to form a stable distribution inside the cavity is  $\sim 1/\kappa$, where $\kappa$ is the coupling loss of the cavity. This means that the optical field can form a stable cavity mode within a time scale ($\sim$ns, as shown below) much less than the coherence time and the spontaneous lifetime ($\sim$ms) of the erbium ions. Therefore, one can use the creation and annihilation operators to describe the dynamics of the cavity modes. The equation of motion for the cavity mode is
\begin{equation} \label{eq.cavity}
\dot{a}=(-i\Delta_c-\frac{\kappa}{2}) a +\sqrt{\kappa}A_{in}+ig P,
\end{equation} 
where $A_{in}$ is the input driving field ( {signal} pulse or control $\pi$ pulses in our case).

Since the cavity frequency $\Delta_c$ can be electrically shifted, according to the Purcell effect \cite{Purcell1946}, emission of {the} atomic ensemble can either be suppressed if $\Delta_c$ is off{-}resonance with the atomic frequency, or enhanced if it is on-resonance. Note also that the DC-stark shift coefficient of erbium ions in LiNbO$_{3}$ is typically 10\,kHz/(Vcm$^{-1}$) \cite{Simon2006}, which means that the ion frequency can also be shifted by $\sim$100\,MHz under the investigated conditions. On one hand, such a shift of the erbium ions is negligible compared to the $\Delta_c$ shift. On the other hand, applying another DC field with opposite polarity can easily cancel the effect of the erbium Stark shift on the memory. Moreover, because the Stark interaction of rare-earth-doped crystals tends to be anisotropic, it is possible to detune the ion frequency in the opposite direction to the cavity frequency, being an advantage due to the enhanced ion-cavity detuning.

Based on the effect of $\Delta_c$, we can both silence the first echo and enhance the second by applying a DC electric field. To get an intuitive picture of the requirements of our scheme, we first consider an input optical pulse with a linewidth much narrower than $\kappa$. In such a case the field inside the cavity can instantaneously follow the variation of the signal field,  i.e., the adiabatic approximation $\dot{a} \approx 0$ applies. Note also that the input-output relation  of a cavity is
\begin{equation} \label{eq.input_output}
A_{\text{out}}+A_{\text{in}} =\sqrt{\kappa}\,a, 
\end{equation} 
At the echo-emission stage when the input $A_{\text{in}}$ is absent, one can write the output of the cavity as
\begin{equation} \label{eq.cav_output}
    A_{\text{out}}= \frac {\sqrt{\kappa}} {i\Delta_c+\frac{\kappa}{2}} i g P. 
\end{equation}
The polarization term  $P$ in Eq.(\ref{eq.cav_output}) originates from the stored collective excitation of the erbium ions. It is clear now if $\Delta_c$ can be detuned to a value much larger than $\kappa$, the output echo can be effectively silenced.

\begin{figure}
\centering
\includegraphics[width=0.45\textwidth]{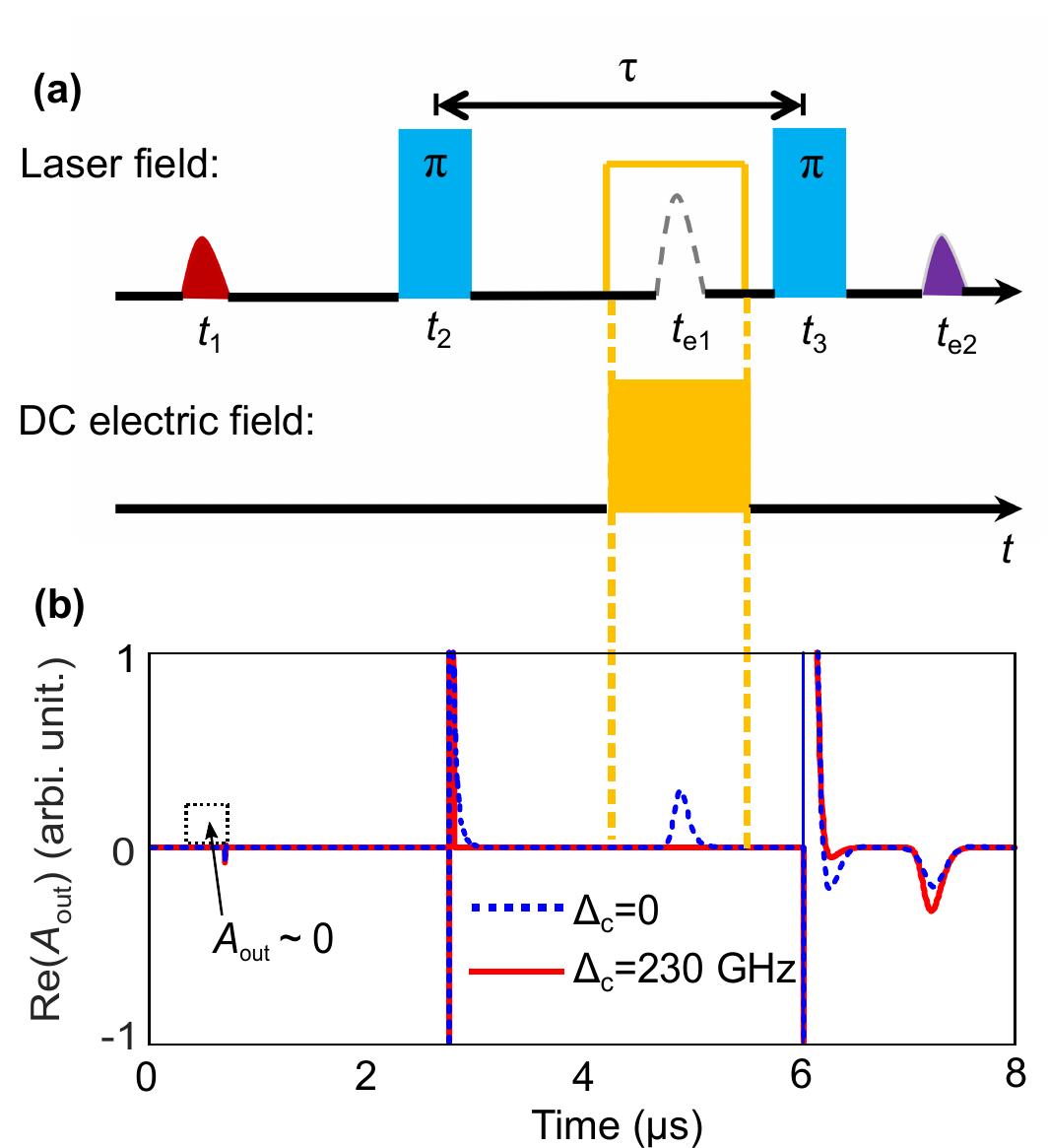}
\caption{Pulse {sequence}. (a) the input pulse at $t_1$ (upper, red), $\pi$ pulses at $t_2$ and $t_3$ (upper, blue), and the external electric field at $t_{e_1}$ (lower, gold). The first echo of the first $\pi$-pulses is silenced by {an} external electric field at $t_{e_1}$. The second echo (upper, purple) is emitted at $t_{e_2}$. (b) The output field of the pulse sequence with (red solid) or without (blue dashed) external electric field to shift the cavity frequency.  }
\label{fig.pulse_seq}
\end{figure}

The sequences of control optical pulses and applied voltage of our scheme are illustrated in Fig. \ref{fig.pulse_seq}(a): 
\begin{itemize}
    \itemsep -1.5em 
    \item [I.]
    At $t=t_{1}$, an optical pulse, which is the signal to be stored, enters the cavity. To obtain high{-}efficiency quantum storage, perfect absorption of the input pulse is needed. For a weak input pulse with few photons (then $\rho_{11}-\rho_{22} \approx 1$ applies), the polarization term $P$ is a linear function of the cavity field: $ig P = - \Gamma a /2$, with $\Gamma$ being a constant. If $\dot{a} \approx 0$ (the cavity mode $a$ can instantaneously follow the change of input pulse), we can obtain from Eq. (\ref{eq.cavity}) and (\ref{eq.input_output}) that 
\begin{equation} \label{eq.Aout_impedance}
A_{\text{out}}(t)=\frac{-i\Delta_c+\frac{\kappa}{2}-\frac{\Gamma}{2}}{i\Delta_c+\frac{\kappa}{2}+\frac{\Gamma}{2} }\,A_{\text{in}}(t).
\end{equation}
When $\Delta_c = 0$, there is impedance-matching $\kappa = \Gamma$ leading to ideal absorption of the input pulse \cite{Moiseev2010, Afzelius2010}. By designing the coupling loss $\kappa$ of the input grating coupler and choosing a proper density of erbium ions {$n_\text{er}$}, the impedance matching condition can be reached so that the input pulse are perfectly absorbed, resulting in  $A_{\text{out}}(t) \approx 0$.\\
    \item[II.]
    At $t=t_{2}$, a rephasing $\pi$-pulse is applied to the erbium ensemble, which reverses the phase of $\rho_{21}(\delta)$.\\
    \item[III.] 
    At $t = t_{e1}$, where $t_{e1} - t_2 = t_2-t_1$, the coherence $\rho_{21}$ with respect to different detuning $\delta$ restores in phase and stimulates an echo. If no voltage is applied, the echo would be emitted into the LiNbO$_{3}$ cavity. As aforementioned{,} this echo is noisy and not suitable for quantum memories. To silence this echo, an external electric field is applied between $t_{2}$ and $t_{3}$, as shown by the DC field sequence in Fig. \ref{fig.pulse_seq}(a). The DC field can shift the cavity frequency $\Delta_c$ out of resonance with the {rephased} collective polarization $P$, suppressing the collective emission as an echo. \\
    \item[IV.]
    At $t= t_3$,  where $t_3-t_2 = \tau$, a second $\pi$-pulse is applied to cancel the phase differences accumulated between $t_3$ and  $t_{e1}$, and, in addition, to bring those ions excited by the first $\pi$ pulse back to their ground states. \\
    \item[V.]
    At $t = t_{e2}$, where $t_{e2} - t_3 = t_3-t_{e1}$, a secondary echo with low noise is released from the cavity.
\end{itemize}
 
To show that efficient storage is experimentally feasible, we calculate the output of the scheme by numerically solving Eq. (\ref{eq.rho12})-(\ref{eq.input_output}). In the present work, the focus is on proposing the principles and mechanisms of a storage scheme, and the building and computation of the realistic model \cite{Johann2018} will be addressed in future work .  The loaded quality-factor of the nano-beam cavity is $Q=9\times 10^4$ (the intrinsic $Q$ in simulations can be as high as $5\times10^6$), and the {mode} volume of which is $V= 0.19 \, \mu m^3$. While erbium ions in bulk LiNbO$_3$ crystal have an inhomogeneous linewidth of $G_{in}=180$ GHz and optical coherence time of $T_{2} = 1.8$\,ms and electric dipole moment of $d_{12} = 3.5 \times 10^{32}$ C$\cdot$m \cite{Thiel2011}, we used the parameters when erbium ions are doped in LiNbO$_{3}$ thin film, which are G$_{in}=166$ GHz and $T_{2} = 0.18$ ms\cite{0Optical}. 

The performance with (red line) and without (blue dashed line) an applied voltage is shown in Fig. \ref{fig.pulse_seq}(b). At $t_{1}$, a  light pulse with {a} duration of {0.16}\,$\mu$s  is input to the cavity. According to a given coupling loss $\kappa$ of our cavity,  we set the concentration of erbium ion as {225} ppm to realize the impedance-matching condition. Perfect absorption is observed in Fig. \ref{fig.pulse_seq}(b), where the output at $t_{1}$ is $A_{\text{out}}(t_{1}) \sim 0$. 

With the absence of an applied DC field, both a primary echo and a secondary echo can be seen after applying $\pi$ pulses, as shown by the blue dashed line in Fig. \ref{fig.pulse_seq}(b). However, if an electrical field is turned on to shift $\Delta_c$$=$230\,GHz out of resonance during the emission period of the first echo, the first echo can be silenced to a negligible level. At the same time the amplitude of the second echo, compared to its counterpart without applied field, is enhanced. These results suggest that one can both silence the unwanted echo and enhance the second by using the electro-optic effect of LiNbO$_{3}$.

\begin{figure}
\centering
\includegraphics[width=0.45\textwidth]{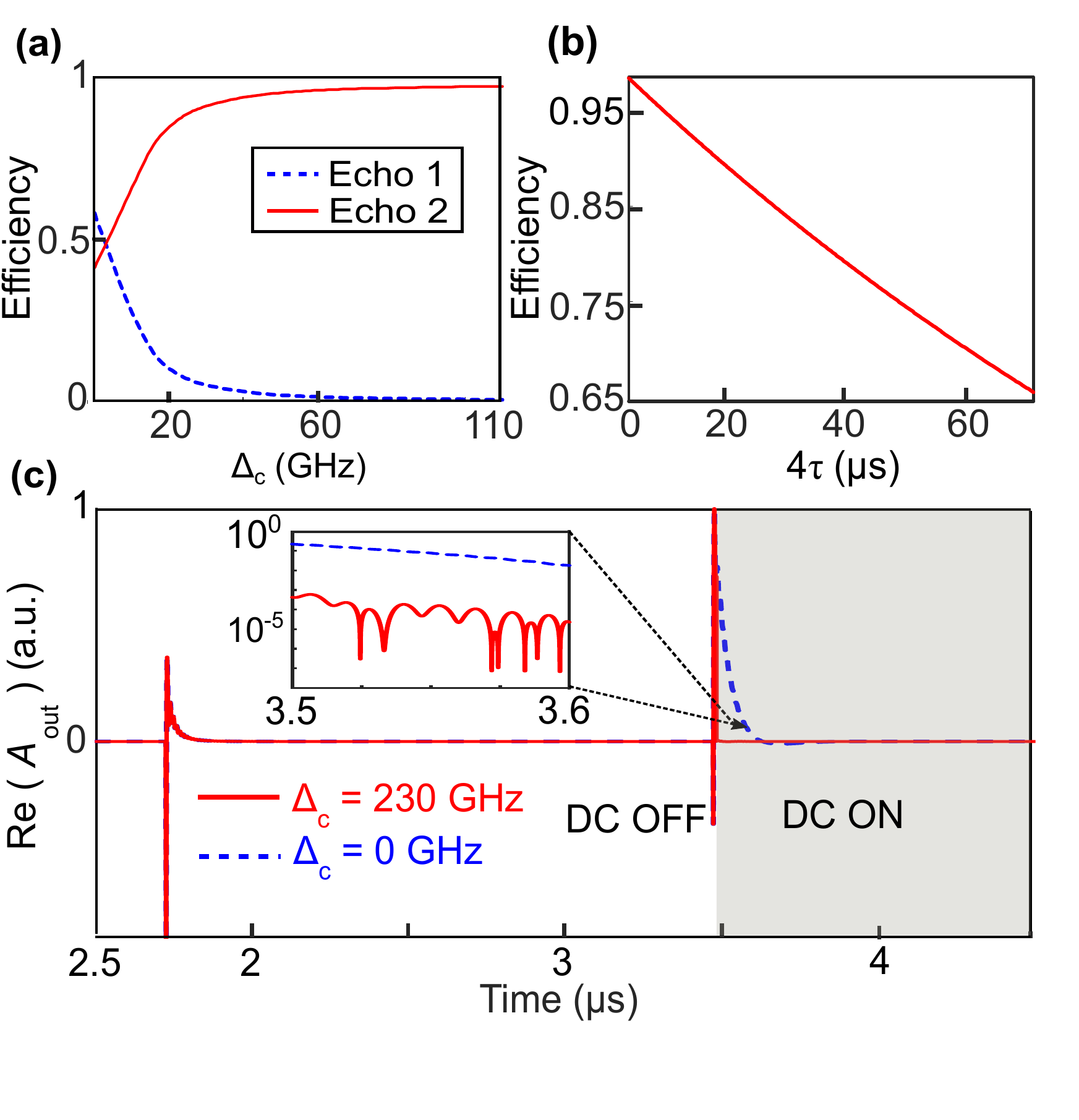}
\caption{Memory efficiency and noise.  (a) Efficiencies of the first echo (dashed blue) and second echo (solid red) as a function of $\Delta_c$. (b) Efficiency as a function of the interval between the signal pulse and the revival echo. The maximum efficiency is limited by $T_2$ of erbium ions, which is 180\,$\mu$s in calculations. $\eta$ depends on $\tau$ through a exponential function {$\exp(-4\tau/T_2)$}. ({c}) Noise analysis. The noise level with and without applied DC electric field are shown in solid {red} and dashed {blue} line. The noise caused by the free induction decay of $\pi$-pulses can be significantly reduced when switching on a large $\Delta_c$ (grey shade). The inset shows the noise ratio of two situations with $\Delta_c=0$ and $\Delta_c=230$\,GHz. The noise can be switched off by $\Delta_c$ whenever needed. }
\label{fig.dependences}
\end{figure}
 
To efficiently cancel the first echo, one needs a large $\Delta_c$  as suggested by Eq (\ref{eq.cav_output}). In Fig. \ref{fig.dependences}(a), we compare the emissions of the two echoes for increasing $\Delta_c$.  When there is no external DC field $\Delta_c =0$, both the first and the second echo are present and of similar weights. With increasing $\Delta_c$, the efficiency of the second echo rises up to {97\%}, accompanied by the decrease of the first one. Note that to obtain a high-contrast ratio between the first and the second echo, $\Delta_c$ is just required to be larger than the cavity linewidth $\kappa$ (2\,GHz in our case), rather than the inhomogeneous linewidth of erbium ions (166\,GHz in our case).  This is evident in Fig. \ref{fig.dependences}(a), where for $\Delta_c \approx 60$\,GHz the ratio between the first and the second echo is already 1.3\%. Therefore a high-Q cavity can reduce the needed DC voltage and facilitate the application of the scheme.

Further increasing $\Delta_c$ above 60\,GHz leads to little increase of the efficiency above $97\%$. This is because the memory operation takes time, during which some ions decohere and cannot be rephased by $\pi$-pulses. The efficiency depends on the decoherence time according to $\eta=\exp(-{4\tau}/{T_2})$, where $4\tau$ is twice the interval between the signal pulse and the revival echo, as shown in Fig. \ref{fig.dependences}(b). In this paper, $T_2 = 180$\,$\mu$s is used from a relevant experiment that measured the coherence time of erbium ions in thin-film LiNbO$_{3}$ waveguides \cite{wang2022}. When rare-earth ions are doped into LiNbO$_{3}$, static defects like charge compensation and compositional disorder cause large inhomogeneous linewidths and can also lead to decoherence \cite{Macfarlane2004}. In addition, the presence of non-zero nuclear spin isotopes $^{93}$Nb, $^7$Li, and $^6$Li causes additional decoherence due to their dynamic magnetic interactions with erbium ions. Note that the $T_2=180$\,$\mu$s in our scheme already enables a theoretical efficiency of 97\%. If the operation time $4\tau$ is reduced, such an efficiency can be extended to meet the requirement of quantum-error-correction applications \cite{Jerry2012}. To further extend the memory storage time without losing high efficiency, one can use the hyperfine ground states of erbium ions. When an erbium isotope with nonzero nuclear spin is used, hyperfine ground states offer the possibility of realizing the Zero-First-Order-Zeeman effect, showing the prospects of extending the coherence times by orders of magnitudes \cite{Rancic2018a, Chen2016c, Rakonjac2020a}.

 
By introducing a tunable LiNbO$_{3}$ cavity, not only can the emission of echoes be controlled, but also the noise of such a memory can be suppressed to a negligible level, as shown in Fig. \ref{fig.dependences}(c). For quantum memories based on photon echoes, typical noise originates from free induction decay or spontaneous emission generated by $\pi$-pulses. Shown in Fig. \ref{fig.dependences}(c) is the output with $\pi$ pulses present but no input signal. Without an applied electric field, the free-induction-decay tails of both $\pi$ pulses immediately arise once the $\pi$ pulses are switched off, as shown by the blue dashed curve of Fig. \ref{fig.dependences}(c). If an electric field is applied after the second $\pi$ pulse to shift the cavity frequency {by} 230\,GHz, both the free induction decay and spontaneous emission (red curve) are suppressed. In our calculations, a ratio of more than 10$^4$ between the noise levels with and without an applied field can be obtained, as shown in the inset of Fig. \ref{fig.dependences}(c).  The low noise thus makes the memory suitable for working at single photon levels. If the input optical state is a one-photon state, then the noise-limited fidelity of the output, according to the calculation method in reference \cite{Jobez2015}, can be more than 99.8\% (depending on the read-out time of the memory).

In our scheme, the primary echo is silenced by $\Delta_c$. This is quite different {from} other protocols such as the Revival-of--Silenced-Echo (ROSE) scheme that utilizes the spacial phase mismatching effect\cite{2011Revival}, the Hybrid-Photon-Echo-Rephasing scheme that utilizes the electric Stark effect\cite{2011Photon}, and the Light-Shift-Photon-Echo-Rephasing scheme that utilizes the light{-}induced frequency shift effect\cite{2016light}. The echo cancellation in all these schemes is the result of a vanishing ensemble polarization due to the destructive interference of ions with different accumulated phases (spatial or temporal). At the moment $t_{e1}$ of our scheme, all ions are already rephased, which means that an ensemble polarization is built up and can otherwise emit photons if $\Delta_c$ is in tune. The silenced echo is based on the Purcell effect rather than the out-of-phase superposition of ions. This difference has brought several advantages, as listed below, to our scheme.

The most important one is that the scheme not only shifts the requirement on controllability from erbium ions to LiNbO$_{3}$, but also bring in the feature of LiNbO$_{3}$. Many memory protocols, such as CRIB and AFC, rely on preparation steps. The preparation not only introduces technical difficulties, but also imposes additional requirements on working ions. For example, in addition to excellent coherence properties of working ions, CRIB also relies on extra electric or magentic field to realize controllable linewidth broadening, which can lead to a compromise between coherence and controllability. The AFC protocol only works for ions with auxiliary long-life-time states. In our scheme, all these requirements are relaxed. When building a memory, one can just focus on utilizing the best experimental conditions to achieve the longest coherence time, as the write-in and read-out of the memory are now controlled by the LiNbO$_{3}$. This means that the protocol can take full advantage of the 1.5\,$\mu$m telecom emission and the long coherence times of erbium ions. Moreover, benefiting from the host LiNbO$_{3}$ thin film, such a device is easy to integrate within modern photonic circuits. The number of memory units can be scaled up with current technology, and the operation of each unit can be independently controlled by electrodes, ideal for integrated quantum chips.

\begin{table}[!]
\renewcommand\arraystretch{1.3}
\caption{\label{tab.single}Theoretical speculation for single-ion quantum memory.}
\begin{ruledtabular}
\begin{tabular}{cccc|c}
Q&V(${\mu m}^3$)&{$n_\text{er}$}$(\text{ppm})$&$N_h$ &Efficiency($\%$)\\
\hline 

$2\times10^5$ & 0.28 & 105 & 3.1
& 97 \\
$3\times10^5$ & 0.3 & 70 & 2.2
& 97 \\
$4\times10^5$ & 0.28 & 50 & 1.5
& 97 \\
$8\times10^5$& 0.26 & 26 & 0.7
& 97 \\
$7\times10^6$ & 0.23 & 3 & 0.07
& 97\\

\end{tabular}
\end{ruledtabular}
\end{table}


The multimode capacity of the memory can be limited by the linewidth of the optical cavity. In the above calculation, the linewidth of the cavity is $\sim 2$\,GHz and the input pulse is $\sim 10$\,MHz, thus the multimode capacity is on the order of $10^2$.  Using a high $Q$ cavity reduces the multimode capacity, but on the other hand, can further enhance the light-matter interaction. For high $Q$ cavities, fewer erbium ions are needed for high optical absorption, which, together with the small mode volume of photonic-crystal cavities, enables us to exploit the possibility of quantum memories based on single ions. The numerical results of calculations on $\eta$ for several combinations of $Q$ and erbium concentrations $n_{\text{er}}$ are shown in Table \ref{tab.single} (note that in these calculations, the mode volume V and the duration of the two $\pi$ pulses are also adjusted to maintain the efficiency $\eta>97\%$). The number of ions that can be resolved in its homogeneous linewidth is approximated by $N_h = n_{\text{er}} V \Gamma_h / \Gamma_{in}$, where $\Gamma_{h} = 1/(\pi T_2) \sim $1\,MHz is chosen for the homogeneous linewidth , and $\Gamma_{in} = 166$\,GHz is the inhomogeneous linewidth \cite{0Optical}. For $Q \approx 10^5$ (LiNbO$_{3}$ micro cavities with $Q \approx 10^8$ has been reported \cite{Wang:20}), the number of interacting erbium ions can easily be reduced to single digits. For erbium doped crystals with Gaussian broadening, working on the edge of the inhomogeneous line can further reduce $N_h$ and facilitate single ion detection. With decreasing erbium concentration, the scheme shows potential for realizing solid-state single-ion quantum memories, where the property of  addressable single ions can be used to predict events of successful storage \cite{2011A} and to establish deterministic entanglement \cite{1996Quantum}.

In summary, we propose an on-chip quantum memory scheme by utilizing the electric-optic effect of LiNbO$_{3}$. Applying a DC electric field can change the refractive index of LiNbO$_{3}$ and the frequency. Therefore, it can change the frequency of a LiNbO$_{3}$ photonic cavity and control the emission of echoes due to the Purcell effect. Our numerical results show that the echo readout efficiency approaches unity. Compared with the technical difficulties in memory schemes, the  electric-field controllability of lithium niobate on a phtonic chip is industrially compatible. 
It has been demonstrated that the erbium ions in a patterned thin-film LiNbO$_{3}$ are able to preserve the coherence properties of bulk material. At high magnetic field and low concentrations, the coherence time can be further extended, which in turn can increase the maximum memory efficiency. Furthermore, when an erbium isotope with nonzero nuclear spin is used, hyperfine interactions offer the possibility of using the ZEFOZ effect\cite{Rancic2018a, Chen2016c, Rakonjac2020a} to extend the coherence times by orders of magnitude. We thus expect it is experimentally feasible to build an integrated quantum memory with high efficiency and long coherence times using Er$^{3+}$:LiNbO$_3$ thin films. The scheme also offers new ideas for exploiting quantum memories, such as noise suppression by electric fields and single-ion quantum memories.

\vline

The authors wish to acknowledge financial support
from the National Natural Science Foundation of China (No. 62105033 and No. 12174026), the Start-up Fund of Beijing Institute of Technology, and the Science and Technology Innovation Project of Beijing Institute of Technology.


\end{document}